\documentclass[reprint, prl, aps, floatfix, nofootinbib, superscriptaddress]{revtex4-2}
\usepackage{amsmath}
\usepackage{rotating}
\usepackage{graphicx}
\usepackage{chemformula}
\usepackage{mathtools}
\usepackage{tabularx}
\usepackage{color}
\usepackage{bm}
\usepackage{dcolumn}
\usepackage{ulem}
\usepackage[colorlinks=true, linkcolor=blue, citecolor=blue, urlcolor=blue]{hyperref}
\newcommand{\gb}[1]{{\color{black}#1}}
\newcommand{\gbb}[1]{{\color{black}#1}}

\newcommand{\fl}[1]{{\color{black}#1}}

\begin{document}

\title{\fl{Breakdown of Universal Whirling Order in a Heisenberg Tsai-Type Approximant}}
%Previous title: Antiferromagnetic orders in Au-Al-Tb and Au-Al-Gd 1/1 approximant crystals studied by x-ray magnetic diffraction}
\author{Farid Labib}
\affiliation{Research Institute for Science and Technology, Tokyo University of Science, Tokyo 125-8585, Japan}
\author{Guillaume Beutier}
\affiliation{Université Grenoble Alpes, CNRS, Grenoble INP, SIMaP, Grenoble, France}
\author{Alessandro Bombardi}
\affiliation{Diamond Light Source, Harwell Science and Innovation Campus, Didcot, Oxfordshire, United Kingdom}
\author{Marc de Boissieu}
\affiliation{Université Grenoble Alpes, CNRS, Grenoble INP, SIMaP, Grenoble, France}
\author{Ryuji Tamura}
\affiliation{Department of Materials Science and Technology, Tokyo University of Science, Tokyo 125-8585, Japan}

\date{\today}

\begin{abstract}
    \fl{Noncoplanar whirling magnetic order has recently been proposed as a universal magnetic state in non-Heisenberg Tsai-type quasicrystal approximants. Here, x-ray resonant magnetic scattering measurements on Heisenberg and non-Heisenberg Au--Al--$R$ ($R$ = Gd, Tb) $1/1$ approximants reveal a clear breakdown of this universality in the Heisenberg limit. While the non-Heisenberg Au--Al--Tb system is well described by the previously identified whirling magnetic structure, the Heisenberg Au--Al--Gd counterpart exhibits qualitatively different azimuthal-angle dependences that are incompatible with the universal whirling-order model. Despite nearly identical crystal structures and comparable magnetic energy scales, both systems stabilize distinct antiferromagnetic ground states with the same propagation vector $k = (1\,0\,0)$, demonstrating that the apparent universality of whirling order is not a consequence of Tsai-cluster geometry alone, but requires spin anisotropy. These results further reveal the emergence of a competing manifold of nearly degenerate magnetic states in the Heisenberg limit.}

\end{abstract}

\maketitle

%\section{Introduction}

\fl{Quasicrystals \cite{Shechtman1984,Shechtman1985} and their periodic counterparts, known as approximant crystals (ACs) \cite{Goldman1993}, offer a distinct playground for exploring magnetic phenomena inaccessible in conventional periodic structures. Numerous examples of long-range ferromagnetic and antiferromagnetic order have been reported not only in ACs of various alloy systems (\cite{nawa2023JPCS,labib2024a,Farid2,Labib2025JACS,Labib2022PRB,Labib2024PRB,Tamura2010}, to name a few) but also in quasicrystals \cite{Tamura2021,Takeuchi2023,Tamura2025}, indicating that magnetic order is not inherently incompatible with these materials and can be controlled and designed based on the currently available guidelines, as described elsewhere \cite{Labib2026MT}.

Despite these advances, the microscopic nature of long-range magnetic order in these materials remains much less understood. Recent neutron diffraction studies on Tsai-type Au–SM–rare-earth (SM = Al, Ga, Si) ferromagnetic and antiferromagnetic $1/1$ ACs containing non-Heisenberg rare-earth moments ($R = \mathrm{Tb}$, $\mathrm{Dy}$, and $\mathrm{Ho}$) \cite{Nawa2023,labib2024a,labib2026, Thilakan2024, Sato2019} have revealed a common noncoplanar ``whirling’’ magnetic structure, represented by the magnetic space groups $Ipm^{\prime}\bar{3}^{\prime}$ and $R\bar{3}$, respectively, despite differences in chemical composition and rare-earth species. Such robustness across non-Heisenberg Tsai-type systems has led to the expectation that the geometry of the Tsai cluster itself may play a dominant role in stabilizing the whirling order.}

Whether the whirling magnetic order is primarily stabilized by the Tsai-cluster geometry alone or instead by other parameters such as rare-earth spin anisotropy remains a topic of ongoing debate. In Gd-based systems, in particular, the nearly vanishing orbital moment suppresses single-ion anisotropy and drives the system toward isotropic Heisenberg behavior. Removing anisotropy is expected to substantially enhance spin configurational freedom, and enlarge the manifold of competing magnetic states as reported in recent Monte Carlo simulations \cite{Sugimoto1,Sugimoto2,Farid2}. Yet, experimental evidence for such expectation has remained inaccessible, largely due to the strong neutron absorption of Gd that severely restricts conventional neutron diffraction studies.

\begin{figure}[!t]
    \centering
    \includegraphics[width=1\linewidth]{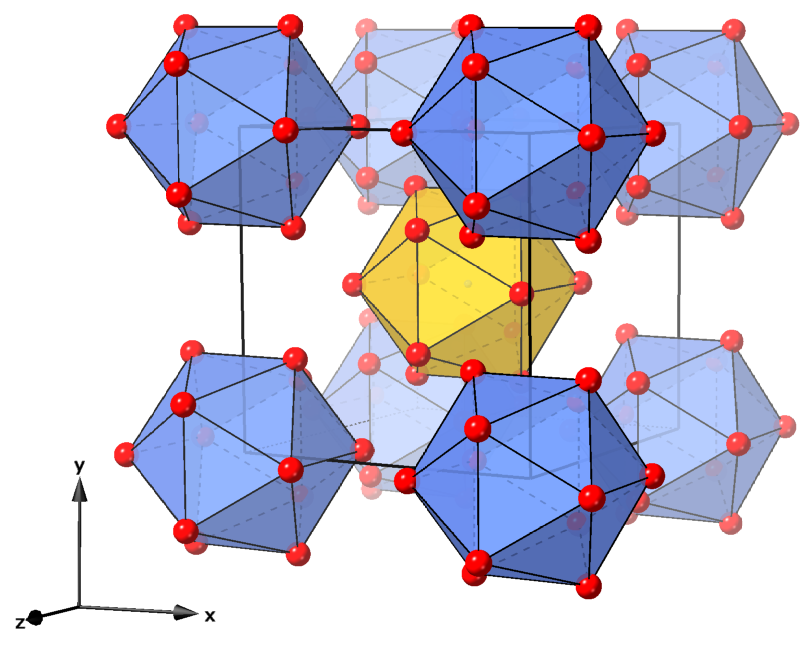}
    \caption{Arrangement of Gd ions on the icosahedron shells in the cubic unit cell of Au--Al--Gd $1/1$ AC  (refer to \cite{Farid2} for structural details). The colors reflect the antiferromagnetic coupling between corner and body-center icosahedra of the cubic lattice, although all icosahedra are structurally identical.}
    \label{fig:bcc-icosahedra-4}
\end{figure}

Here, X-ray resonant magnetic scattering (XRMS) \cite{Paolasini} measurements are performed on single crystals of antiferromagnetic Au–Al–Tb and Au–Al–Gd $1/1$ ACs, as representative examples of non-Heisenberg and Heisenberg Tsai-type spin systems, respectively. The two compounds share identical crystal structures crystalizing in the cubic symmetry (space group  $\mathit{Im}\bar{3}$), with the Tb and Gd ions fully occupying the $24g$ Wyckoff position (see FIG.~\ref{fig:bcc-icosahedra-4} for the configuration of $R$ atoms on the icosahedron shell derived from a refined atomic structure in a representative Au--Al--Gd $1/1$ AC with a lattice parameter of 14.81~\AA \cite{Farid2}). Despite their structural similarity, the Au–Al–Tb and Au–Al–Gd $1/1$ ACs differ fundamentally in their spin anisotropy: the former is strongly anisotropic, whereas the latter behaves as a nearly isotropic Heisenberg system. To elucidate the influence of spin anisotropy on the magnetic ground state, XRMS was employed. The technique is particularly well suited for this purpose because of the strong enhancement of magnetic scattering at the rare-earth $L_{2,3}$ absorption edges \cite{DoonGibbs1988,Kim2005}, making it an effective probe of magnetic structures in strongly neutron-absorbing rare-earth compounds.

The present measurements reveal a clear breakdown of the proposed universality of the whirling magnetic order in the Heisenberg limit. While the Au–Al–Tb 1/1 AC is well described by the previously established whirling magnetic order, the Au–Al–Gd counterpart exhibits qualitatively distinct magnetic scattering behavior incompatible with existing models. These findings provide the first experimental evidence that unlike common expectations, cluster geometry alone is insufficient to uniquely determine magnetic order in Tsai-type ACs and instead identify spin anisotropy as a decisive parameter governing the magnetic-state selection.

\begin{figure}[!t]
    \centering
    \includegraphics[width=\linewidth]{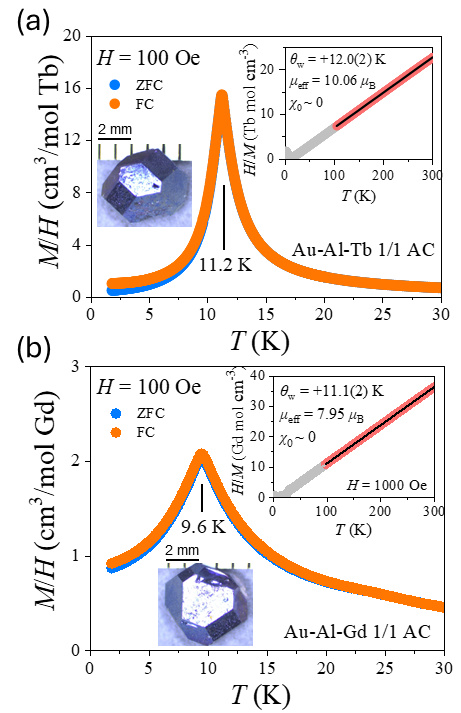}
    \caption{Low temperature magnetic susceptibility $M/H$ of (a)Au--Al--Tb, and (b) Au--Al--Gd $1/1$ AC single crystals under $H = 100$~Oe. The insets in (a) and (b) show corresponding inverse magnetic susceptibilities ($H/M$) measured under $H = 1000$~Oe as well as the optical microscopu images of the corresponding single crystals.}
    \label{fig:figure3}
\end{figure}

\fl{Single crystals of Tsai-type Au--Al--Tb and Au--Al--Gd $1/1$ ACs with average diameters of $\sim$5 mm (see the bottom-left inset of FIG.~\ref{fig:figure3}) were grown by a solution growth method following a protocol described elsewhere \cite{Farid2}. Bulk magnetic properties were measured using a SQUID magnetometer (Quantum Design MPMS3). The temperature dependence of $M/H$ under an applied field of $H=100$ Oe (FIG.~\ref{fig:figure3}) exhibits sharp cusps in both field-cooled (FC) and zero-field-cooled (ZFC) curves at $T_{\mathrm N}=11.4$ K and $9.4$ K for Au--Al--Tb and Au--Al--Gd $1/1$ ACs, respectively, indicating the onset of antiferromagnetic order. Measurements along the [100], [110], and [111] directions revealed no detectable magnetic anisotropy. The corresponding high-temperature inverse susceptibilities ($H/M$), measured under $H=1000$ Oe (upper-right insets of FIG.~\ref{fig:figure3}), were fitted in the range $100<T<300$ K using the Curie--Weiss expression $\chi(T)=N_{\mathrm A}\mu_{\mathrm{eff}}^2\mu_{\mathrm B}^2/[3k_{\mathrm B}(T-\theta_{\mathrm W})]+\chi_0$. The fits yield $\theta_{\mathrm W}=+12.0$ K and $+11.1$ K for Au--Al--Tb and Au--Al--Gd, respectively, indicating comparable magnetic energy scales despite their fundamentally different spin anisotropies. The extracted effective moments, $10.06~\mu_{\mathrm B}$ and $7.95~\mu_{\mathrm B}$, are close to the free-ion values of $9.74~\mu_{\mathrm B}$ and $7.94~\mu_{\mathrm B}$ expected for $\mathrm{Tb}^{3+}$ and $\mathrm{Gd}^{3+}$, respectively \cite{Blundell2001}, confirming well-localized $\mathrm{R}^{3+}$ moments. Together with their nearly identical crystal structures, the similarities in magnetic energy scale establish Au--Al--Tb and Au--Al--Gd 1/1 ACs as ideal controlled platform for isolating the role of spin anisotropy in magnetic-state selection.}

Next, XRMS measurements were performed at beamline I16 of Diamond Light Source \cite{Collins} with photon energies tuned to the Tb $L_3$ edge (7.514 keV) and Gd $L_3$ edge (7.243 keV) \gbb{and polarisation of the incident beam linear and perpendicular to the scattering plane}. Azimuthal-angle scans were carried out by rotating the sample around the scattering vector of selected magnetic reflections. In the present geometry, the magnetic scattering intensity scales as \gbb{$|\mathbf{k'(\psi)}\!\cdot\!\mathbf{M}(hkl)|^2$} \cite{Hill}, allowing different projections of the magnetic structure factor $\mathbf{M}(hkl)$ \gbb{onto the scattered beam direction $\mathbf{k'(\psi)}$} to be probed as a function of azimuthal angle \gbb{$\psi$ (See Figure S3 in the Supplemental Material for a description of the scattering geometry \cite{SM})}. Integrated intensities were obtained from rocking-curve scans after background subtraction and self-absorption correction. In both compounds, clear magnetic Bragg peaks are observed at charge-forbidden reflections satisfying $h + k + l = 2n + 1$, demonstrating the breaking of the body-centered translational symmetry, which provides unambiguous evidence for the establishment of long-range antiferromagnetic order. The sharpness of these magnetic reflections indicates well-developed long-range magnetic correlations, consistent with previous XRMS studies on Tsai-type \ch{Cd$_6$Tb} and \ch{Cd$_6$Ho} $1/1$ approximant crystals \cite{Kim,Kreyssig}.

\begin{figure}[!b]
    \centering
    \includegraphics[width=\linewidth]{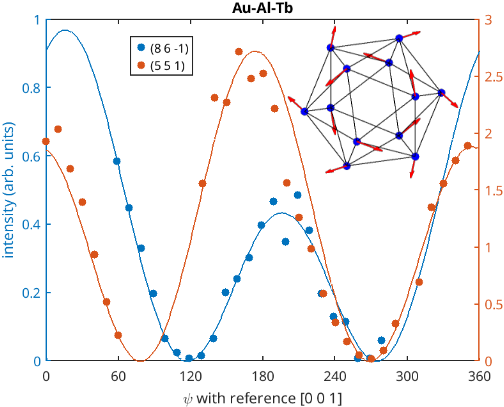}
    %\caption{(a) Azimuthal scan of the $(5,5,1)$ magnetic reflection of Au--Al--Tb, together with the model corresponding to the whirling magnetic order proposed in Ref.~\cite{Sato2019}. (b) Spin configuration on an icosahedron based on the whirling model with $\phi = 118^\circ$ (moments tilted by $86^\circ$ from the fivefold axis).}
    \caption{Azimuthal scans of the $(8,6,-1)$ and $(5,5,1)$ magnetic reflections of Au--Al--Tb. Dots: experimental; plain curves: models corresponding to the whirling magnetic order proposed in Ref.~\cite{Sato2019} and depicted in inset (view from a $[1,1,1]$ axis): the spins are tilted by $86^\circ$ from the fivefold axes.}
    \label{fig:figure4}
\end{figure}

In the case of Au--Al--Tb $1/1$ AC, %azimuthal-angle scans were measured for several magnetic reflections, including $(5\,5\,1)$ and $(8\,6\,-1)$. 
\gb{azimuthal scans were measured for the $551$ and $86\bar{1}$ magnetic reflections.}
The resulting azimuthal dependencies %(see FIG.~\ref{fig:figure4}(a) for a representative $(5\,5\,1)$ reflection) 
\gb{(FIG.~\ref{fig:figure4})} display characteristic modulations encoding the directions of $\mathbf{M}(551)$ and $\mathbf{M}(86\bar{1})$.
%display pronounced periodic modulations characteristic of noncoplanar magnetic order (\textcolor{red}{Refer to supplementary material for azimuthal dependences of other reflections}). 
%To interpret these data, the analysis was carried out within the symmetry framework of the $\mathit{Im}\bar{3}$ space group. Within this symmetry, the magnetic structure is described by irreducible representation~2 (irrep~2), which allows noncoplanar spin configurations parameterized by a single free angle $\phi$, which defines the orientation of the magnetic moments within the $(x,y)$ mirror plane with respect to the crystallographic $x$ axis. This degree of freedom generates a family of noncoplanar magnetic structures compatible with the crystallographic symmetry.
%\fl{In the case of the Au--Al--Tb $1/1$ AC}, 
\gb{The data are well reproduced by the magnetic structure proposed by Sato \textit{et al}~\cite{Sato2019}: it corresponds to 
%an irreducible representation of the $\mathit{Im}\bar{3}$ space group 
the magnetic space group $I_p m'\bar3'$ (204.5, OG setting)
for which the moments are in the local mirror plane of the crystallographic position, at \gbb{$\phi=$28$^\circ$} from one of the cubic axes, thereby at an angle of 86$^\circ$ from the five-fold axis of the site. The magnetic configuration of one icosahedron is shown in inset of FIG.~\ref{fig:figure4}. Attempts to refine the value of the single free parameter, \textit{i.e.} the orientation $\phi$ of the moments in the mirror plane, did not lead to significant improvement. %for both reflections. 
Importantly, this magnetic structure admits only antiphase domains, to which these measurements are not sensitive.} 
%Using the specific value $\phi = 118^\circ$, corresponding to the whirling magnetic structure previously proposed for the Au$_{72}$Al$_{14}$Tb$_{14}$ $1/1$ AC \cite{Sato2019}, the calculated azimuthal-angle dependences reproduce both the phase and overall symmetry of the experimental data (Fig.~\ref{fig:figure4}(a)). The resulting real-space spin configuration (Fig.~\ref{fig:figure4}(b)) consists of Tb moments tilted by approximately $86^\circ$ away from the local fivefold axes, forming a noncoplanar whirling structure.
%Although the overall agreement supports the realization of a whirling magnetic order in Au--Al--Tb 1/1 AC, attempts to refine the parameter $\phi$ independently for different reflections yield inconsistent values. 
% OUTDATED \gb{The deviation between the data and the fit, in particular regarding the position of the minima, suggests that the magnetic structure might not be entirely captured by a single irreducible representation.}
%This suggests that the magnetic structure may slightly deviate from the ideal irrep~2 model or involve additional degrees of freedom not captured within the minimal symmetry description. 
%Nevertheless, 
The present results firmly confirm noncoplanar whirling antiferromagnetic order in the Au--Al--Tb 1/1 AC stabilized by strong single-ion anisotropy, consistent with \cite{Sato2019}.

\gb{In the case of the Au--Al--Gd $1/1$ AC, a large number of magnetic reflections were also found at structurally-forbidden $h+k+l=2n+1$ reflections. They are purely resonant at the Gd $L_3$ edge and their temperature dependence is consistent with a second-order phase transition with a Néel temperature of $\sim9~K$ (see Supplemental Material \cite{SM}). Similarly to the case of the Au--Al--Tb 1/1 AC, they correspond to the commensurate propagation vector $k = (1\,0\,0)$, indicating antiferromagnetic coupling between corner and center clusters of the BCC lattice (see FIG.~\ref{fig:bcc-icosahedra-4}). However, their azimuthal-angle dependence }
%The azimuthal-angle dependencies observed in the Au--Al--Gd $1/1$ AC 
differ qualitatively from those measured in the Au--Al--Tb counterpart. 
%Azimuthal scans were recorded for at least six magnetic reflections from different crystallographic facets, including $(5\,4\,0)$, $(5\,6\,0)$, and $(0\,1\,8)$ 

\begin{figure}[!b]
    \centering
    \includegraphics[width=\linewidth]{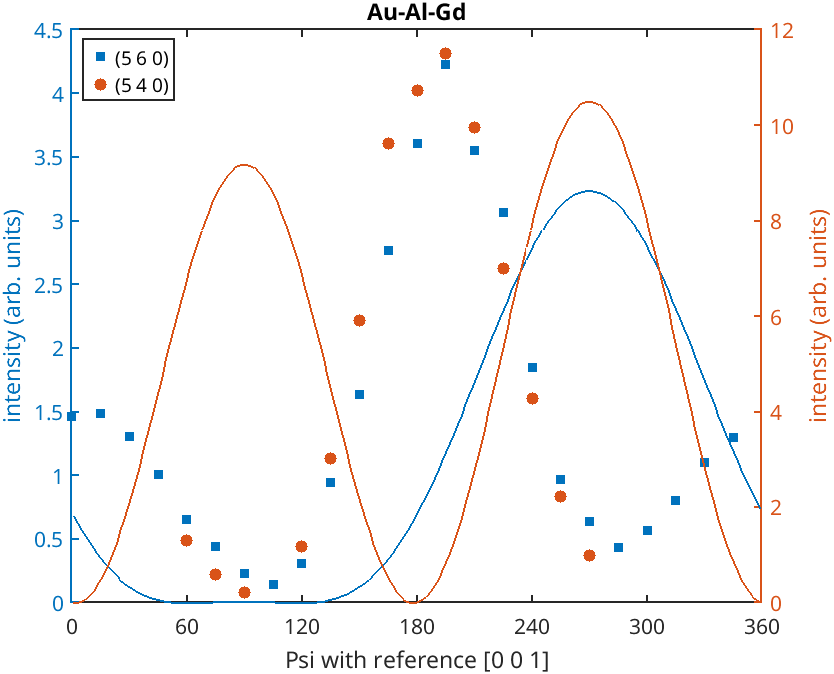}
    \caption{Azimuthal scan of the $(5,6,0)$ as $(5,4,0)$ magnetic reflections of Au--Al--Gd. Dots: experimental; plain curves: models corresponding to the whirling magnetic order proposed in Ref.~\cite{Sato2019} (see also inset of FIG.~\ref{fig:figure4}). }
    \label{fig:figure5}
\end{figure}

\gb{Partial azimuthal scans were recorded for six magnetic reflections: $(5\,4\,0)$, $(5\,6\,0)$, $(7\,7\,7)$, $(0\,0\,13)$, $(1\,0\,8)$, and $(0\,1\,8)$. The first three were measured from the same $(1\,1\,0)$ facet and the last three from the same $(0\,0\,1)$ facet. The azimuthal dependence of the $(5\,4\,0)$ and $(5\,6\,0)$ reflections are shown in FIG.~\ref{fig:figure5}, and the other ones in the Supplemental Material \cite{SM}.}
%(\textcolor{red}{see Supplemental Fig... for the complete set of azimuthal dependencies}). 
%All observed magnetic reflections can be indexed by a commensurate propagation vector $k = (1\,0\,0)$, unambiguously indicating the commensurate nature of the antiferromagnetic order in Au--Al--Gd $1/1$ AC. 
However, none of the measured azimuthal profiles can be reproduced by the whirling magnetic model, irrespective of the choice of the free parameter $\phi$ within the same magnetic space group. 
This discrepancy is systematic and persists across all measured reflections, providing compelling evidence that the whirling magnetic order seen in non-Heisenberg Tsai-type $1/1$ ACs cannot describe the magnetic ground state of the Au--Al--Gd counterpart.
%To clarify the origin of such incompatibility, 
A broad range of alternative magnetic models was examined, including other symmetry-allowed irreducible representations of the $\mathit{Im}\bar{3}$ space group, as well as the Heisenberg models proposed in Refs.~\cite{Sugimoto2,Suzuki2021,Sugimoto1}, and the models of Ref.~\cite{Watanabe2021}. 
\gbb{A triple-$k$ model inspired by the magnetic structure recently reported for {Cd$_6$Tb} \cite{Woodland} was also tested.}
While some of the models can reproduce the azimuthal dependence of individual reflections, none provides a consistent description of all measured reflections. %within the commensurate $\boldsymbol{k}=(100)$ framework. In particular, irrep~1 predicts systematic extinctions of $(0\,0\,2n+1)$ magnetic reflections, in direct contradiction with experimental observations. 
\gbb{More details about the models and the fits are provided in the Supplemental Material \cite{SM}.}

\gb{An important unknown is the occurrence of magnetic domains, if allowed by the magnetic structure. Most models break at least one symmetry of the $\mathit{Im}\bar{3}$ space group, hence allowing the occurrence of domains related by the lost symmetries. Azimuthal scans can thus reflect the contributions of several domains with unknown weights. Nevertheless, such consideration still does not allow to single out a model that would provide a significantly better fit than the others. As mentioned above, the symmetry of the magnetic structure of the Au--Al--Tb ACs allows only antiphase domains, which eases the analysis as we cannot distinguish them in this experiment. In the case of Heisenberg models, the magnetic texture could even vary continuously at the microscale, as they are energetically invariant by an arbitrary and global rotation of the spins \cite{Sugimoto2}.}

A striking feature of the Au--Al--Gd data is the near-identical azimuthal dependence of the $(5\,4\,0)$ and $(5\,6\,0)$ reflections (FIG.~\ref{fig:figure5}), despite their distinct crystallographic indices. Furthermore, the azimuthal profile of the $(5\,6\,0)$ reflection remains essentially unchanged upon crossing the Néel temperature \gb{(see Supplemental Material \cite{SM})}. This behavior suggests the selection of a unique magnetic variant or a near-single-domain magnetic state %associated with the commensurate propagation vector $\boldsymbol{k}=(100)$, 
rather than a superposition of multiple symmetry-related domains. Such domain selection is highly nontrivial in a system with nearly isotropic Heisenberg spins and contrasts sharply with the anisotropy-driven domain selection observed in non-Heisenberg Tsai-type compounds. 

\fl{The contrasting behavior of Au–Al–Tb and Au–Al–Gd 1/1 ACs carries an important implication. Given that the two compounds share nearly identical crystal structures, the same propagation vector, and comparable magnetic energy scales, the contrast directly identifies spin anisotropy as the missing control parameter behind the previously proposed universality of whirling order. Moreover, the present XRMS results substantially narrow the allowed magnetic-structure space in isotropic Tsai-type systems and provide direct experimental evidence for a breakdown of the previously proposed universality of whirling order in the Heisenberg limit.}

\fl{In summary, x-ray resonant magnetic scattering measurements on single crystals of Tsai-type Au--Al--Tb and Au--Al--Gd $1/1$ ACs reveal distinct antiferromagnetic ground states despite their nearly identical crystal structures and comparable magnetic energy scales. While both systems stabilize commensurate order with propagation vector $k=(1\,0\,0)$, only the non-Heisenberg Au--Al--Tb compound could be described by the universal noncoplanar whirling magnetic structure. In contrast, the Heisenberg Au--Al--Gd counterpart exhibits magnetic behavior incompatible with existing whirling or symmetry-based models. These findings establish a clear breakdown of the proposed universality of whirling order in the Heisenberg limit and identify spin anisotropy as an essential selector of magnetic topology. More broadly, they suggest that suppressing anisotropy destabilizes the proposed universal whirling state and promotes a competing manifold of nearly degenerate magnetic states, providing a new perspective on magnetic-state selection in geometrically frustrated cluster magnets.}

%In summary, x-ray resonant magnetic scattering measurements on single crystals of Tsai-type Au--Al--Tb and Au--Al--Gd $1/1$ approximant crystals have revealed distinct antiferromagnetic orders, despite their closely related crystal structures and magnetic energy scales. \gb{Both crystals stabilize a commensurate antiferromagnetic order characterized by the propagation vector $k = (1\,0\,0)$, meaning antiferromagnetic coupling between the corner and center icosahedral clusters of the BCC lattice, but the detailed magnetic structures of the icosahedra differ strongly between both crystals.} The non-Heisenberg Au--Al--Tb compound hosts a noncoplanar whirling magnetic structure driven by strong single-ion anisotropy, whereas the Heisenberg Au--Al--Gd compound develops %a commensurate antiferromagnetic order characterized by the propagation vector $k = (1\,0\,0)$
%a magnetic order that cannot be explained by existing whirling or symmetry-based models. \gb{An unambiguous determination of the magnetic structure remains out of reach with the existing data, but may be obtained with XRMS measurements on a field-cooled crystal, in order to favor a uniform magnetic texture at the mesoscale. Nevertheless,} the results demonstrate that spin anisotropy plays a decisive role in selecting the magnetic ground state of Tsai-type approximants. 

\begin{acknowledgments}
\emph{Acknowledgments} --- 
\gb{This work was carried out with the support of Diamond Light Source, instrument I16 (proposal MM-33633).}
This work is supported by the Japan Society for the Promotion of Science through Grants-in-Aid for Scientific Research (Grants No. JP19H05817, No. JP19H05818, No. JP21H01044, and No. JP24K17016) and the Japan Science and Technology agency, CREST, Japan, through a grant No. JPMJCR22O3. 
\end{acknowledgments}

\bibliography{references.bib}

\end{document}